# $Sr_4Al_2O_7$: A New Sacrificial Layer with High Water Dissolution Rate for the Synthesis of Freestanding Oxide Membranes

Leyan Nian,† Haoying Sun,† Zhichao Wang,† Duo Xu, Hao Bo, Shengjun, Yan, Yueying Li, Jian Zhou, Yu Deng, Yufeng Hao,* and Yuefeng Nie*

National Laboratory of Solid State Microstructures
Jiangsu Key Laboratory of Artificial Functional Materials
College of Engineering and Applied Sciences
Collaborative Innovation Center of Advanced Microstructures
Nanjing University
Nanjing 210023, P. R. China
† These authors contributed equally to this work.

*Electronic address: haoyufeng@nju.edu.cn
*Electronic address: ynie@nju.edu.cn

**Abstract**: Freestanding perovskite oxide membranes have drawn great attention recently since they offer exceptional structural tunability and stacking ability, providing new opportunities in fundamental research and potential device applications in silicon-based semiconductor technology. Among different types of sacrificial layers, the (Ca, Sr, Ba)$_3$Al$_2$O$_6$ compounds are most widely used since they can be dissolved in water and prepare high-quality perovskite oxide membranes with clean and sharp surfaces and interfaces. However, the typical transfer process takes a long time (up to hours) in obtaining millimeter-size freestanding membranes, let alone realize wafer-scale samples with high yield. Here, we introduce a new member of the SrO-Al$_2$O$_3$ family, Sr$_4$Al$_2$O$_7$, and demonstrate its high dissolution rate, about 10 times higher than that of Sr$_3$Al$_2$O$_6$. The high-dissolution-rate of Sr$_4$Al$_2$O$_7$ is most likely related to the more discrete Al-O networks and higher concentration of water-soluble Sr-O species in this compound. Our work significantly facilitates the preparation of freestanding membranes and sheds light on the integration of multifunctional perovskite oxides in practical electronic devices.

**Keywords**:
sacrificial layer, Sr$_4$Al$_2$O$_7$, freestanding oxide membranes, molecular beam epitaxy

**Introduction**:
　　The unique properties of freestanding perovskite oxide membranes, such as extraordinary strain tunability,[1-3] stacking ability and declamping effect,[4-7] facilitate the ferroelectric and ferromagnetic phase engineering,[1-3, 8] super-elasticity[9-12] and functionality integration of perovskite oxides on silicon wafer,[6, 13-17] *etc*. The recent surge of research interest in these freestanding membranes is driven by the advances of sacrificial layers for selective wet etching method.[18-22] Especially, the seminal work by Lu *et al.* used water-soluble $Sr_3Al_2O_6$ as the sacrificial layer in synthesizing perovskite oxide membranes with high crystalline quality and sharp surfaces.[20] The lattice of $Sr_3Al_2O_6$ (cubic, $a$ = 15.844 Å) closely matches four unit cells (u.c.) of $SrTiO_3$ (STO) ($a_{STO}$ = 3.905 Å, 4 × $a_{STO}$ = 15.62 Å), the prototype perovskite oxide, allowing their epitaxial growth on perovskite oxide substrates and preparation a variety of freestanding oxide membranes.[6, 20, 23, 24] Moreover, by varying the ratio of Ca:Sr:Ba in (Ca, Sr, Ba)$_3Al_2O_6$, the lattice constants of the sacrificial layer can be adjusted for a wide range (15.263 Å to 16.498 Å),[20, 25, 26] further facilitating the synthesis of coherently-strained epitaxial films by reducing the possibility of lattice relaxation.[27, 28] Up to now, many other oxides, such as $La_xSr_{1-x}MnO_3$,[19, 29] $SrRuO_3$,[13, 22] $SrVO_3$,[30] and $SrCoO_{2.5}$,[31] have also been demonstrated as sacrificial layers to release membranes from the bottom substrates. Nonetheless, the (Ca, Sr, Ba)$_3Al_2O_6$ compounds remain to be the most widely used sacrificial layers these days on account of the high-quality products and their facile dissolution in water, an environmentally friendly solution, at room temperature.

One of the great interests on these perovskite oxide membranes is the integration of their rich fascinating functionalities, including ferroelectricity, ferromagnetism, dielectric properties, and others, with silicon-based semiconductor technologies.[32, 33] For example, high-density switchable skyrmion-like polar nanodomains, oxide ferroelectric tunnel junction memories, and high-κ perovskite oxides have already been integrated with silicon and demonstrate excellent properties.[14, 16, 34] However, such a wet-etching process using $Sr_3Al_2O_6$ is typically time-consuming, taking several hours or even days for millimeter-size samples, which hinders its potential in practical electronic applications. Moreover, minimizing the possible detrimental effects induced by water-soluble process is an important consideration since many oxide films are not very stable in water or even in air.[13, 35-38] For example, the metastable state of the infinite-layer nickelate superconducting films with $Ni^{1+}$ valance state becomes volatile when staying in air or moisture atmosphere.[36, 39, 40] Therefore, exploring a new sacrificial layer with an ultrafast water solubility rate to avoid possible dissolving damage is of great importance in future electronic applications as well as fundamental science research.

In this work, we synthesize a new water-soluble compound of the $SrO-Al_2O_3$ system, $Sr_4Al_2O_7$, by molecular beam epitaxy and demonstrate its high water dissolution rate for rapid fabrication of perovskite oxide membranes. The precise flux ratio of $Sr_4Al_2O_7$ is controlled by reflection high-energy electron diffraction (RHEED) and its high crystallinity and microscopic atomic structure are confirmed by X-ray diffraction (XRD) and scanning transmission electron microscopy (STEM). The water dissolution rate of $Sr_4Al_2O_7$ is extremely high, almost 10 times higher than that of $Sr_3Al_2O_6$. Our work broadens the options of water sacrificial layers and provides an ultrafast and effective route for the integration of functional perovskite oxide membranes with silicon or other semiconductors.

**Results and Discussion**:

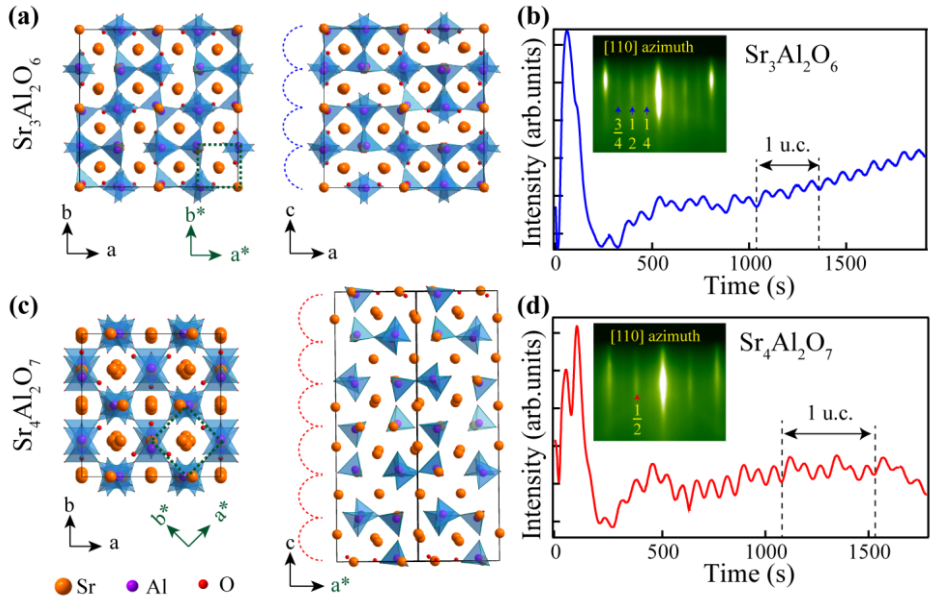

Figure 1. Schematics of the crystal structure of $Sr_3Al_2O_6$ (a) and $Sr_4Al_2O_7$ (c). The green coordinates are for STO lattice direction. The black coordinates are for $Sr_3Al_2O_6$ or $Sr_4Al_2O_7$. (b, d) Time-dependent RHEED intensity oscillations and patterns of $Sr_3Al_2O_6$ (b) and $Sr_4Al_2O_7$ (d) films grown on STO substrates.

In order to increase the water solubility of the sacrificial later, introducing more water-soluble ingredients in the $SrO-Al_2O_3$ binary compound is highly desired. As pointed out by Lu *et al*., an important reason why $Sr_3Al_2O_6$ is water-soluble is that its Al-O network is discrete rather than a fully connected framework.[20] Further reducing the connections between Al-O clusters would enhance the water solubility of this class of compounds. In fact, another important aspect accounting for it water solubility is the hydrotropic Sr-O networks which is interlaced with the discrete Al-O clusters. It is well-known that all alkaline earth metal oxides (except BeO) readily react with water to form metal hydroxide salts, $MO(s)+H_2O(l) \rightarrow M(OH)_2(s)$ (M:alkaline earth metal).[41] Even in a more complex compound, the Sr-O species could also be dissolved in water, so that boiling water can remove the surface SrO layer and form $TiO_2$-terminated $SrTiO_3$ substrates.[42, 43] Therefore, increasing the concentration of *M*-O (*M*:alkaline earth metal) is also a key ingredient to enhance the water solubility of the $SrO-Al_2O_3$ binary compound. Following these two guidelines, $M_4Al_2O_7$ is an ideal compound since its Al-O network is more discrete and contains higher concentration of *M*-O than $M_3Al_2O_6$ (Figure 1(a) and 1(c)). $M_4Al_2O_7$ is reported to have a orthorhombic phase (space group:Cmca).[44] In the past, the studies on $M_4Al_2O_7$ were mainly focused on luminescent properties of polycrystalline samples.[45, 46] The epitaxial growth of single crystalline $M_4Al_2O_7$ films and their water solubility have not been systematically explored up to date.

To synthesize high-quality $Sr_4Al_2O_7$ films, we firstly calibrated the Sr and Al fluxes by growing $Sr_3Al_2O_6$ film on STO substrate, since the RHEED diffraction patterns and intensity oscillations of $Sr_3Al_2O_6$ exhibit a strong dependence on the Sr/Al flux ratio as reported in our previous work.[47] Following the calibration process, we optimize the Sr/Al flux to be 3:2 during the co-deposition growth of $Sr_3Al_2O_6$ film. Next, the Sr/Al flux ratio is calibrated to be 2:1 by increasing the Sr temperature source according to the thermodynamic relationship between vapor pressure and temperature.[47, 48] After the flux calibration, $Sr_4Al_2O_7$ epitaxial films were grown on STO substrate using a co-deposition growth mode. During film growth, the crystalline quality of $Sr_3Al_2O_6$ and $Sr_4Al_2O_7$ films was monitored by *in situ* RHEED. As shown in the inset in Figure 1(b), the RHEED pattern of $Sr_3Al_2O_6$ surface taken along the [110] azimuth of STO substrate shows a four-fold reconstructed pattern since its lattice constant is 4 times larger than STO ($a_{STO} = 1/4 \times a_{Sr3Al2O6}$).[20, 47] Interestingly, the RHEED pattern taken with the same condition of $Sr_4Al_2O_7$ shows a two-fold reconstructed pattern (inset in Figure 1(d)), indicating its in-plane periodicity may be different from $Sr_3Al_2O_6$. Compared with $Sr_3Al_2O_6$, the calculated $Sr_4Al_2O_7$ has smaller *a* and *b* lattice constants of 10.80 Å and 11.22 Å, respectively. Rotating the in-plane lattice of $Sr_4Al_2O_7$ by 45° gives a better lattice match with STO substrate (average of -0.34 %), compared with the $Sr_3Al_2O_6$ (+1.41 %). This in-plane lattice orientation is further supported by STEM results discussed later. Moreover, the RHEED intensity oscillations also exhibit different features. The RHEED oscillations of $Sr_3Al_2O_6$ exhibit a four periods of intensity oscillations in the growth of one u.c., which is consistant with four familiar atomic sublayers along [001] direction in one u.c. (Figure 1 (a)). In contrast, the RHEED oscillations of $Sr_4Al_2O_7$ show a different pattern with three peaks per period (Figure 1(c) and (d)), indicating more atomic sublayers per unit cell in $Sr_4Al_2O_7$ compared to $Sr_3Al_2O_6$. This is consistent with our expectation of 6 and 4 atomic sublayers in $Sr_4Al_2O_7$ and $Sr_3Al_2O_6$, respectively (Figure 1(a) and (c)).

The crystalline quality of $Sr_4Al_2O_7$/STO and $Sr_3Al_2O_6$/STO samples was measured by X-ray diffraction (XRD). In Figure 2(a), the XRD $2\theta$-$\omega$ scan of 10.2 nm $Sr_4Al_2O_7$ and 9.5 nm $Sr_3Al_2O_6$ sandwiched between 10 u.c. STO film and STO substrate shows much larger out-of-plane lattice constants of $Sr_4Al_2O_7$ film than $Sr_3Al_2O_6$. The reciprocal space mappings of these films indicate that the films have the same in-plane lattice constant as the STO substrate without any lattice relaxation (Figure 2(c) and (e)). Based on the XRD data, the lattice constant *c* of $Sr_4Al_2O_7$ is estimated to be 25.47 Å by the Bragg formula, which is much larger than $Sr_3Al_2O_6$ (*c* =15.87 Å). These results are consistent with our first-principle density functional theory (DFT) calculations, 25.78 Å and 15.81 Å for $Sr_4Al_2O_7$ and $Sr_3Al_2O_6$ bulk states, respectively. The rocking curve measurements were performed to investigate the crystal quality of these films ($Sr_3Al_2O_6$ and $Sr_4Al_2O_7$) and STO substrate. As shown in Figure 2(b), a relatively small full width at half maximum (FWHM) values of 0.04 ° and 0.03 ° for the (00$\underline{12}$) peak of the $Sr_4Al_2O_7$ film and (002) peak of STO substrate, respectively. The comparable FWHM value of films with substrate demonstrates that the crystalline quality of $Sr_4Al_2O_7$ film is basically limited by the underlying STO substrate. Similarly, the $Sr_3Al_2O_6$/STO heterostructure also shows narrow rocking curves and high crystalline quality. Moreover, the $2\theta$-$\omega$ curves show clear intensity oscillations at small angles indicating the atomically smooth surface of these films, consistent with the AFM images shown in Figures 2 (d) and (f). By fitting the Kiessig fringes data using Diffrac Leptos (Bruker) software,[49] the film thicknesses of $Sr_4Al_2O_7$ and $Sr_3Al_2O_6$ are extracted to be 9.8 nm and 9.1 nm, which well agrees with RHEED monitored thickness of 4 u.c. and 6 u.c., respectively.

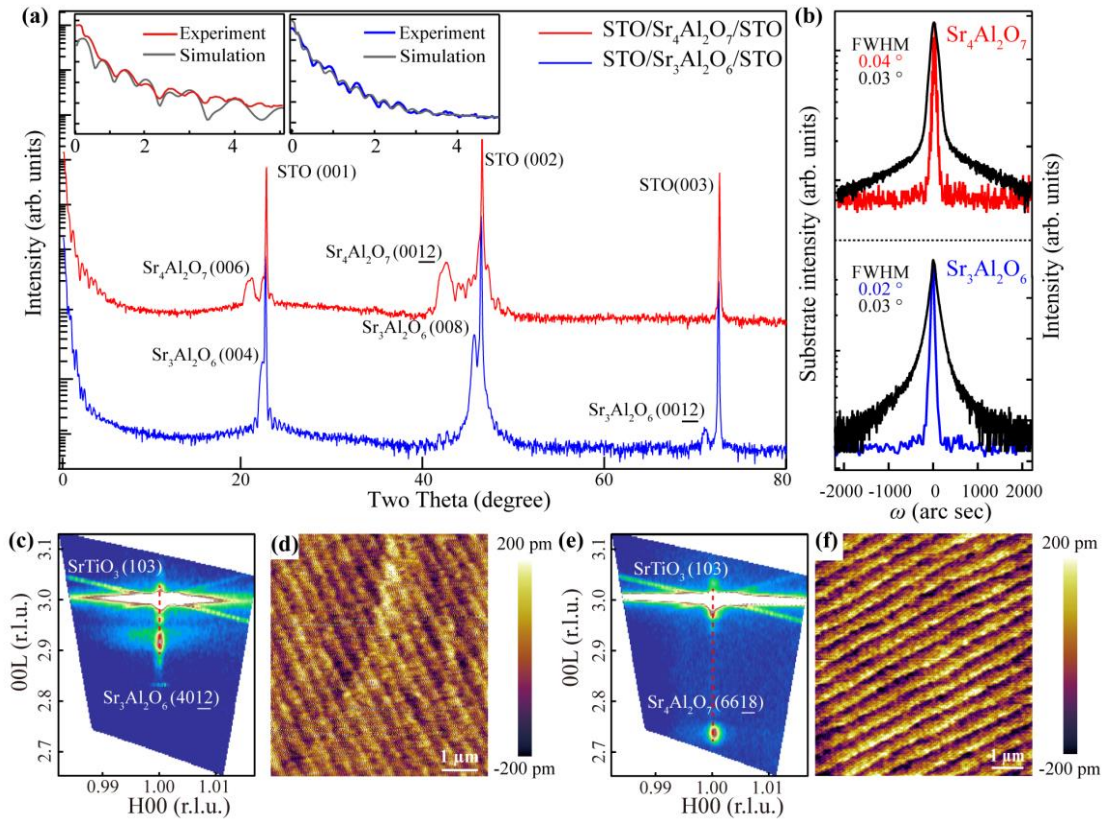

Figure 2. (a) 2θ-ω scans of 10 u.c. STO/10.2 nm $Sr_4Al_2O_7$/STO and 10 u.c. STO/9.5 nm $Sr_3Al_2O_6$/STO heterostructures. The insets show the enlarged view around 0-5 degrees of the two samples and their Kiessig fringes simulated curves. (b) Rocking curves of (00$\underline{12}$) diffraction of $Sr_4Al_2O_7$ film (red), (008) diffraction $Sr_3Al_2O_6$ film (blue) and (002) diffraction of STO (black), respectively. (c, e) Reciprocal space mapping (RSM) around the STO (103) reflection for 10 u.c. STO/10.2 nm $Sr_4Al_2O_7$/STO (c), and 10 u.c. STO/9.5 nm $Sr_3Al_2O_6$/STO samples (e). The red dashed line is shown as a guide to the eye for the fully strained $Sr_4Al_2O_7$ (or $Sr_3Al_2O_6$) film with respect to STO substrate. (d, f) Corresponding AFM topography images of 10 u.c. STO/10.2 nm $Sr_4Al_2O_7$/STO(d), and 10 u.c. STO/9.5 nm $Sr_3Al_2O_6$/STO sample (f).

Cross-sectional STEM images were taken along the [010] axis of STO to examine the microstructural information of these two compounds. The zoom-in scanning transmission electron microscopy high-angle annular dark field (STEM-HAADF) (Figure 3(a) and (c)) and corresponding fast Fourier transform (FFT) images (Figure 3(b) and (d)) of $Sr_4Al_2O_7$ and $Sr_3Al_2O_6$ show the different periodic arrangements, indicating the distinct crystal structures. Since the contrast of different atoms in STEM-HAADF image is approximately proportional to $Z^2$ (Z: atomic number),[50] only heavier Sr and Al atoms are visible while lighter O atoms are invisible. The STEM-HAADF image of $Sr_3Al_2O_6$ is consistent with previous reports.[20] As shown in Figure

3(c), the atomic arrangement of $Sr_4Al_2O_7$ perfectly matches the expected crystal structure when the [110] direction of $Sr_4Al_2O_7$ is aligned with the [100] direction of the STO substrate. Note that the brighter *A*-sites in the unit cell of $Sr_4Al_2O_7$ are occupied by heavier Sr atoms while the slightly darker *B*-sites are co-occupied by both Sr and lighter Al atoms, and thus it manifests the perovskite-like structure. In contrast, the *B*-sites of $Sr_3Al_2O_6$ unit cell is alternatively occupied with pure Al atoms and the mixed Sr/Al atoms, demonstrates alternating light and dark diffraction spots. The differences in lattice arrangement mentioned above is consistent with the FFT images of these STEM-HAADF images.

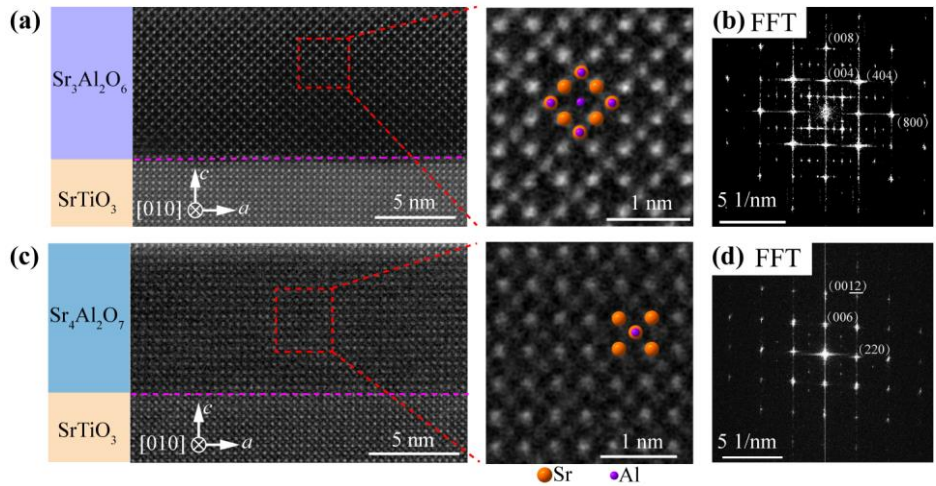

Figure 3. (a, c) Cross-sectional STEM-HAADF images of $Sr_4Al_2O_7$/STO and $Sr_3Al_2O_6$/STO heterostructures. (b, d) The corresponding FFT patterns of (a) and (c).

In order to compare the water dissolution rates of $Sr_4Al_2O_7$ and $Sr_3Al_2O_6$, we synthesized $Sr_3Al_2O_6$ and $Sr_4Al_2O_7$ with same thickness sandwiched between STO film and STO substrates. For a fair comparison, $Sr_3Al_2O_6$, $Sr_4Al_2O_7$, and top STO films are kept the same thickness and all substrates have the same size about 5×5 mm$^2$. A PDMS tape is attached to the top surface to provide a mechanical support for the film, which helps clearly visualize the dissolution process as the sacrificial layer dissolves from the edges to the center (Figure 4(a)). As summarized in Figure 4(b), the dissolution rate of 20 nm (8 u.c.) $Sr_4Al_2O_7$ is about ten times faster than that of 19 nm (12 u.c.) $Sr_3Al_2O_6$. It only takes 6 minutes for $Sr_4Al_2O_7$ to completely dissolve $Sr_4Al_2O_7$ to fabricate a 5×5 mm$^2$ freestanding film., which can greatly reduce the transferring time compared to $Sr_3Al_2O_6$. Note that the slight difference of the film thicknesses is because we synthesized the films with a complete number of unit cells. We also explore the impact of the thickness of $Sr_4Al_2O_7$ and $Sr_3Al_2O_6$ on their dissolution rates (Figure 4(c)). It is clear that the thicker the sacrificial layer, the faster it dissolves, since a larger space between the top film and substrate can allow water to penetrate through more easily. For all thicknesses, the dissolution rates of $Sr_4Al_2O_7$ films are much faster (about 10 times) than that of $Sr_3Al_2O_6$.

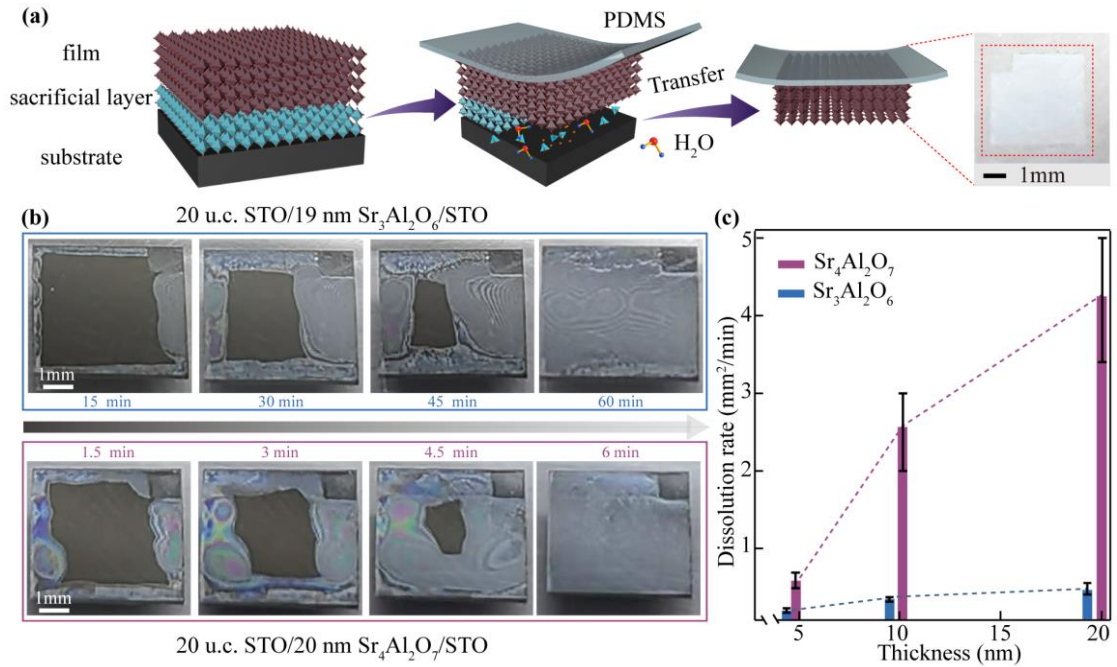

Figure 4. (a) Schematic diagrams of fabricating freestanding STO heterostructures on a PDMS tape. (b) Dissolution process of 20 u.c. STO/19 nm $Sr_3Al_2O_6$/STO and 20 u.c. STO/20 nm $Sr_4Al_2O_7$/STO heterostructures. The sample size is 5×5 mm$^2$. (c) Comparison of the thickness-dependent dissolution rates of $Sr_3Al_2O_6$ and $Sr_4Al_2O_7$.

**Conclusion**: In conclusion, we introduce a new member of the $SrO$-$Al_2O_3$ family, $Sr_4Al_2O_7$, as an effective sacrificial layer for fabricating freestanding perovskite oxide membranes. The dissolution rate of $Sr_4Al_2O_7$ is about 10 times larger than that of $Sr_3Al_2O_6$, which is most likely related to the more discrete Al-O networks and higher concentration of water-soluble SrO in this compound. Also, the dissolution rate strongly depends on the thickness of the sacrificial layer, the thicker the layer the faster the rate. In addition to $Sr_4Al_2O_7$, other members of $M_4Al_2O_7$ ($M$=Ca, Sr, Ba) are also expected to be water-soluble. Similar to the (Ca, Sr, Ba)$_3Al_2O_6$ family, varying the Ca:Sr:Ba ratio in (Ca, Sr, Ba)$_4Al_2O_7$, its lattice constants can be tuned to yield the best lattice-matched buffer layers for the epitaxial growth of different perovskites oxides. This new $M_4Al_2O_7$ compound can greatly improve the efficiency in transferring freestanding membranes, which is important for synthesizing those moisture-sensitive films and sheds light on the integration of fascinating perovskite oxides in practical electronic devices.

**Experimental Section**:
*Film growth*: All the samples were grown by a DCA R450 MBE system. Before growth, commercial STO substrates (10 × 10 × 1 mm$^3$) were etched in buffered 8% HF (NH$_4$F) acid for 65 s and then annealed in oxygen at 1000 °C for 80 min to obtain a TiO$_2$-terminated step-and-terrace surface for atomically-precise film growth. The initial calibration of Sr/Al flux ratio was realized by a quartz crystal microbalance (QCM) and adjusted by changing the Sr (or Al) source temperature to obtain an approximate flux to be 3:2 ($Sr_3Al_2O_6$) or 2:1 ($Sr_4Al_2O_7$). $Sr_4Al_2O_7$ and $Sr_3Al_2O_6$ water-soluble layers were grown on the (001)-oriented $SrTiO_3$ substrates. And the optimal growth condition was found to be $T_{substrate}$ = 700 °C (measured by a pyrometer) at an oxygen partial pressure $P_{oxygen}$ = 1.0 × 10$^{-6}$ Torr (measured by a residual gas analyzer). Subsequently, the $Sr_4Al_2O_7$ and $Sr_3Al_2O_6$ sacrificial layers were covered by a STO layer grown under the same growth condition. A reflection high-energy electron diffraction (RHEED) with 15 kV acceleration voltage was employed to monitor the thickness and crystal quality of as-grown films precisely.

*Structural Characterizations*: After films growth, the STO/$Sr_4Al_2O_7$/STO and $BaTiO_3$/$Sr_4Al_2O_7$/STO heterostructures were cooled to room temperature (~30 °C) under the deposition oxygen partial pressure with a rate of 15 °C min$^{-1}$ and then were taken out. The surface morphologies were characterized by AFM (MFP-3D SA, Asylum). The diffraction peaks were characterized by high-resolution XRD using a Bruker D8 Discover diffractometer. Atomic-scale microstructure of films was revealed using a double-spherical aberration-corrected STEM/TEM FEI Titan G2 60-300 microscope at 300 kV.

*DFT Method*:
The crystal structures of $Sr_3Al_2O_6$, $Sr_4Al_2O_7$, and STO are calculated by the density functional theory in the generalized gradient approximation implemented in the Vienna ab initio simulation package code,[51] in which the projected augmented wave method[52, 53] and the Perdew-Burke-Ernzerhof revised for solids exchange-correlation[54] are used. The plane-wave cutoff energy is 520 eV throughout the calculations. The k-mesh is 2×2×2, 4×4×2, and 9×9×9 for the $Sr_3Al_2O_6$, $Sr_4Al_2O_7$, and $SrTiO_3$, respectively. The crystal structures of the three materials are fully optimized until the maximal residual Hellmann-Feynman forces are less than 0.02 eV/Å. The elastic constants are calculated with the assistance of the Elastic code.[55]

Note: During the manuscript preparation, we noticed that Lingfei Wang *et al.* from University of Science and Technology of China also indenpendtly synthesized and observed similar exciting properties in these $Sr_4Al_2O_7$ compounds

**Acknowledgments**: Leyan Nian, Haoying Sun, and Zhichao Wang contributed equally to this work. This work was supported by the National Key R&D Program of China (2022YFA1402502 and 2021YFA1400400), the National Natural Science Foundation of China (11861161004 and 11974163), and the Fundamental Research Funds for the Central Universities (Grant 0213-14380221).